\RequirePackage{fix-cm}
\documentclass[smallextended]{svjour3}       
\smartqed  
\usepackage{graphicx}
\usepackage{amssymb}
\usepackage{amsmath}

\newcommand{\ket}[1]{| #1 \rangle} 
\newcommand{\bra}[1]{\langle #1 |} 
\newcommand{\braket}[2]{\langle #1 \vphantom{#2} | #2 \vphantom{#1} \rangle} 
\begin{document}

\title{Telecloning of qudits via partially entangled states
}


\author{Gabriel Araneda       \and
        Nataly Cisternas \and
        Aldo Delgado 
}


\institute{Gabriel Araneda \at
              Institut f\"{u}r Experimentalphysik,Universit\"{a}t Innsbruck, Technikerstrasse 25, 6020 Innsbruck, Austria
              \\ \email{gabriel.araneda-machuca@uibk.ac.at}           
           \and
            Nataly Cisternas \at
            Van der Waals-Zeeman Institute for Experimental Physics, Universiteit van Amsterdam, Science Park 904,1098 XH Amsterdam, The Netherlands
      	   \and
		   Aldo Delgado \at
	      Center for Optics and Photonics, Universidad de Concepci\'{o}n, Casilla 4012, Concepci\'{o}n, Chile\\ MSI-Nucleus on Advanced Optics, Universidad de Concepci\'on, Casilla 160-C, Concepci\'on, Chile  \\ Departamento de F\'isica, Universidad de Concepci\'on, Casilla 160-C, Concepci\'on, Chile          }

\date{}

\maketitle

\begin{abstract}
We  study the process of quantum telecloning of $d$-dimensional pure quantum states using  partially entangled pure states as  quantum channel. This process efficiently mixes   optimal universal symmetric cloning with quantum teleportation. It is shown that it is possible to implement  universal symmetric  telecloning in a probabilistic way using unambiguous state discrimination and quantum state separation schemes. It is also shown  that other strategies, such as minimum error discrimination, lead to a  decrease in the fidelity of the copies and that certain partially entangled pure states with maximal Schmidt rank lead to an average telecloning fidelity which is always above the optimal fidelity of measuring and preparation  of quantum states. We also discuss  the case of partially entangled pure states with non-maximal Schmidt rank.  The results presented here are valid  for arbitrary numbers of copies of a single input qudit state of any dimension. 

\end{abstract}

\section{Introduction}
\label{SECTION-I}

The encoding of information in quantum systems exhibits features which do not have classical counterparts, including quantum teleportation \cite{Bennett,Vaidman} and the no-cloning theorem \cite{Wootters}. Quantum teleportation enables quantum states to be transmitted from a sender, Alice, to a receiver, Bob, without sending a physical quantum system. The no-cloning theorem rules out the perfect cloning of unknown quantum states by the linearity of quantum mechanical transformations. It does allow approximate deterministic cloning \cite{Buzek,Werner,Keyl,Alber} and perfect probabilistic cloning \cite{Duan,DelgadoA,DelgadoB}. Approximate deterministic cloning is the creation of imperfect copies of an unknown quantum state with the maximal fidelity permitted by quantum mechanics. Perfect probabilistic cloning is the creation of exact copies of an unknown quantum state in a probabilistic fashion.

Quantum teleportation and approximate universal cloning can be combined to allow a sender to distribute approximate clones to many receiving parties. The sender may first generate all the required clones before teleporting them to the receivers. This strategy requires the use of as many bipartite maximally entangled states as there are clones to be broadcast, as well as the transmission of classical information to complete the teleportation stage. A new and more efficient approach, termed telecloning \cite{Murao1,Murao2}, allows simultaneous conveyance of all the clones by means of a single local measurement carried out by the sender, who publicly broadcasts the result of the measurement. Each receiving party then performs a local quantum operation which depends on the publicly broadcast measurement result and thereafter the approximate clones are finally broadcast. This kind of scheme has been experimentally demonstrated using entangled photons \cite{Chiuri}, where three nonperfect but optimal copies were transmitted using maximally entangled bipartite pure states. This strategy is more efficient than the simple scheme described above since it requires only $O(log_2M)$ maximally entangled bipartite pure states, where $M$ is the number of clones to be distributed.

Telecloning is based on the generation of a multipartite entangled state which is distributed among the parties participating in the telecloning process. In realistic situations, however, generation and distribution do not, in general, lead to a maximally entangled state. Here we study the performance of telecloning when implemented with a partially entangled pure state as a quantum channel. We focus on the errors introduced in the telecloning process and the degradation of the fidelity of the clones. We show that the problem of recovering high fidelity in the copies reduces to the problem of discrimination between non-orthogonal quantum states. We also propose a probabilistic and conclusive method to correct the errors caused by the non-maximal entanglement of the multipartite entangled state, using standards strategies of quantum state discrimination. This article is organised as follows: In section \ref{SECTION-II} we briefly review the telecloning process. In section \ref{SECTION-III} we show how the problem of telecloning via a partially entangled pure quantum state can be related to the problem of discrimination of non-orthogonal quantum states. In section \ref{SECTION-IV} we consider the case of a pure quantum channel of full Schmidt rank, and we combine unambiguous state discrimination with the telecloning scheme to correct the errors introduced by the non-maximal entanglement of the multi-partite entangled state. We show that the fidelity of this process, averaged over the Hilbert space of the states to be cloned, is higher than the fidelity of estimating quantum states for certain quantum channels. Also in this section we study the combination with other discrimination schemes, namely minimum error and quantum state separation. In section \ref{SECTION-V} we comment the case of a pure quantum channel with non-maximal Schmidt rank and our results are summarised in section \ref{SECTION-VI}.

\section{Telecloning process}
\label{SECTION-II}

The linearity of quantum operations forbids the perfect cloning of unknown quantum states. Nevertheless, it is possible to consider an approximate cloning process which creates clones of unknown quantum states with the highest fidelity allowed by quantum mechanics. This optimal process is carried out by applying separate unitary transformations onto a set of quantum systems. We consider a universal and symmetric telecloning process \cite{Murao1,Murao2}. It is universal in the sense that the fidelities of the clones do not depend on the particular state undergoing the cloning process, and it is symmetric in that all of the clones have the same fidelity with respect to the input state. Additionally, each of the copies and the system to be cloned may be spatially separated.

Let us consider  the case of creating $M$ clones from a single copy of an  arbitrary qudit of dimension $d$,  denoted as $1 \rightarrow M$ telecloning \cite{Murao2}. The maximal fidelity for this kind of process is given by the optimal fidelity of the $1\rightarrow M$ universal cloning, $F_{opt}^{1\rightarrow M}=(2M+d-1)/(M+Md)$. The input state to be teleported is the qudit state $\ket{\psi}_X=\sum_{j=0}^{d-1}\alpha_j\ket{j}_X$, where the set of states $\{\ket{j}\}$ is the computational basis for the system $X$. The output state is represented through the basis $\ket{\phi_j}$, which describes $N_0=2M-1$ $d$-dimensional systems, where $M-1$ of them are ancillary systems and the remaining $M$ systems encode the clones. The elements of this basis are written  in terms  of the basis of normalised and symmetrised states $\{\ket{\xi_k^M}\}$. They are given given by
\begin{equation}
\ket{\phi_j}_{AC}=\frac{\sqrt{d}}{\sqrt{d[M]}}\sum\limits\limits_{k=0}^{d[M]}\mathbf{}_{P}\braket{j}{\xi_{k}^{M}}_{PA}\otimes\ket{\xi_{k}^{M}}_{C},
\end{equation}
where $d[M]=(d+M-1)!/M!(d-1)!$. The index $P$ represents the port qudit, which belongs to Alice, $A$ represents the $M-1$ ancillas and $C$ the $M$ qudits which host the clones.  A constructive procedure for the states $\{ \ket{\xi_{k}^{M}}_{C} \}$ is detailed in \cite{Murao2}. The optimal quantum channel for the telecloning process is  given by 
\begin{equation}
\ket{\xi}_{PAC} = \frac{1}{\sqrt{d}}\sum\limits\limits_{j=0}^{d-1}\ket{j}_{P}\otimes\ket{\phi_{j}}_{AC} \label{canal}, 
\end{equation}
which is a maximally entangled state between the port qudit and the $AC$ system. 
Therefore, the joint state of the total system (input state and quantum channel) is given by
\begin{equation}
|\psi\rangle_{XPAC} =|\psi\rangle_{X}\otimes|\xi\rangle_{PAC}.
\end{equation}
This state can be cast as
\begin{align}
\ket{\psi}_{XPAC}=&\sum\limits\limits_{n,m=0}^{d-1}\ket{\Phi_{nm}}_{XP}\frac{1}{\sqrt{d}}\sum\limits\limits_{j=0}^{d-1}\omega^{jn}\alpha_{j}\ket{\phi_{j\oplus m}}_{AC}, \label{sys1}
\end{align}
where $\omega=\exp(2\pi i/d)$ and the generalised Bell states $\{\ket{\Phi_{nm}}\}$ ($n,m=0,\ldots,d-1$) are given by
\begin{equation}
\ket{\Phi_{nm}}_{XP}=\frac{1}{\sqrt{d}}\sum\limits\limits_{k=0}^{d-1}\omega^{kn}\ket{k}_{X}\ket{k\oplus m}_{P},
\label{bell} 
\end{equation}
with $k\oplus m=(j+m)~mod(d)$. Equivalently, the states of the separable basis of systems $X$ and $P$ are given in terms of the Bell states by the expression
\begin{equation}
\ket{k}_X\ket{k\oplus m}_P=\frac{1}{\sqrt{d}}\sum_{n=0}^{d-1}\omega^{-nk}\ket{\Phi_{nm}}_{XP}.
\end{equation}
We now perform a measurement on the  systems $X$ and $P$ in the $\{\ket{\Phi}_{nm}\}$ basis. The outcomes of this measurement are the eigenvalues associated with the indexes $n$ and $m$, and they are transmitted by means of classical communications to the carriers of systems $A$ and $C$. In order to recover the optimal copies, the carriers of these systems apply   local reconstruction unitary operations,  conditioned to the outcomes of the measurement. These operations are  given by 
\begin{align}
U_{nm}^{A} & =  \sum\limits\limits_{j=0}^{d-1}\omega^{-jn}\ket{j}\otimes\bra{j\oplus m},\\
U_{nm}^{C} & =  \sum\limits\limits_{j=0}^{d-1}\omega^{jn}\ket{j}\otimes\bra{j\oplus m},
\end{align} 
for the ancillas and copy systems, respectively. Thereby, the state in the $AC$ system is $\ket{\psi}_{AC}=\sum_{k=0}^{d-1}\alpha_j \ket{\phi_j}_{AC}$, which contains  optimal copies of the input state in each system $C$. The fidelity of each copy is optimal and given by $F_{opt}^{1\rightarrow M}$. 

\section{Telecloning via partial entanglement}
\label{SECTION-III}

Now let us suppose that the quantum channel given by Eq. (\ref{canal}) is not a maximally entangled state between the port qudit and the $AC$ system, and  instead is given by the state
\begin{equation}
\ket{\tilde{\xi}}_{PAC}=\sum\limits\limits_{j=0}^{d-1}c_{j}\ket{j}_{P}\otimes\ket{\phi_{j}}_{AC}\label{partialst},
\end{equation}
where $\{c_j\}$ are $d$  coefficients which define the channel and satisfy $\sum_{j=0}^{d-1}|c_{j}|^{2}=1$. We assume that the channel is written in its Schmidt decomposition so that all the $c_j$ are real, positive numbers.  In this way, the state of the total system is given by
\begin{equation}
\ket{\psi}_{XPAC}  =  \ket{\psi}_{X}\otimes\ket{\tilde{\xi}}_{PAC}, 
\end{equation}
which can be cast as
\begin{equation}
\ket{\psi}_{XPAC}=\sum_{n,m=0}^{d-1}\ket{\Phi_{nm}}_{XP}\sum\limits\limits_{j=0}^{d-1}\alpha_{j}c_{j+m}\exp(-2\pi ijn/d)\ket{\phi_{j+m}}_{AC}.
\end{equation}
If we perform a measurement in the generalised Bell basis Eq. (\ref{bell}), the state of the system is projected onto
\begin{equation}
|\psi\rangle_{nm}^{AC}=\frac{1}{\sqrt{P_{m}}}\sum\limits\limits_{j=0}^{d-1}\alpha_{j}c_{j+m}\exp(-2\pi ijn/d)\ket{\phi_{j+m}}_{AC}\label{simsim},
\end{equation}
where $m$ and $n$ denote the generalised Bell state associated with the measurement result, and $P_{m}$ is the probability of projecting onto the $|\Phi\rangle_{nm}$ state. This probability   depends only on $m$,
\begin{equation}
P_{m} =  \sum\limits\limits_{j=0}^{d-1}|\alpha_{j}|^{2}|c_{j+m}|^{2}\label{pm}.
\end{equation}
Since the values of $n$ and $m$ are known after this measurement, it is possible to apply the corresponding local  reconstruction unitary operations  $U_{nm}^{A}$
and $U_{nm}^{C}$ on the ancillas and copies systems. After this operations the state of the system is described by
\begin{equation}
|\psi'\rangle_{nm}^{AC}=\frac{1}{\sqrt{P_{m}}}\sum\limits\limits_{j=0}^{d-1}\alpha_{j}c_{j+m}\ket{\phi_{j}}_{AC}
,\end{equation}
and the associated density matrix can be written as
\begin{equation}
\rho_{nm}^{AC}=\frac{1}{\sqrt{P_{m}}}\sum\limits\limits_{j,j'=0}^{d-1}\alpha_{j}c_{j+m}\alpha_{j'}c_{j'+m}\ket{\phi_{j}}_{AC}\bra{\phi_{j}}_{AC}.
\end{equation}

Let us study the fidelity  in the simplest case in which two qudit copies ($M=2$) are created. In this case the vectors $|\phi_j\rangle_{AC}$ can be written explicitly  as
\begin{equation}
|\phi_{j}\rangle_{AC_{1}C_{2}}=\sqrt{\frac{1}{2(d+1)}}\sum\limits_{k=0}^{d-1}|k\rangle_{A}\left(|jk\rangle_{C_{1}C_{2}}+|kj\rangle_{C_{1}C_{2}}\right)\label{1-2}.
\end{equation}
The reduced density matrix of one of the copies, e.g. $C_2$, is then
\begin{align}
\rho_{nm}^{C_{2}}  = & \frac{1}{2(d+1)}\sum\limits\limits_{l=0}^{d-1}\ket{l}\bra{l}\nonumber \\
  & +\frac{(2+d)}{2(d+1)}\frac{1}{P_m}\sum\limits\limits_{j,j'=0}^{d-1}\left[\alpha_{j}c_{j+m}\alpha_{j'}c_{j'+m}\right]\ket{j}\bra{j'}.
 \label{RhoreducedPE}
\end{align}
Hence, the local fidelity of the clones $F_{m}^{1\rightarrow2, PE}=\langle\psi|\rho_{nm}^{C_{2}}|\psi\rangle$ with respect to the initial input state for a partially entangled channel is given by
\begin{equation}
F_{m}^{1\rightarrow 2, PE}=\frac{1}{2(d+1)}+\frac{(2+d)}{2(d+1)}\frac{1}{P_{m}}\left(\sum_{k=0}^{d-1}|\alpha_{k}|^{2}c_{k+m}\right)^{2}.
\end{equation}
The average fidelity of the process for a fixed channel is the sum of all the possible fidelities weighted by the probability of obtaining  the different results in the generalised Bell measurement, that is $F^{1\rightarrow 2, PE}=\sum_{m=0}^{d-1}P_{m}F_{m}^{1\rightarrow 2, PE}$, which take the explicit form
\begin{equation}
F^{1\rightarrow 2, PE}=  \frac{1}{2(d+1)}+\frac{(2+d)}{2(d+1)}\sum\limits_{m=0}^{d-1}\left(\sum_{k=0}^{d-1}|\alpha_{k}|^{2}c_{k+m}\right)^{2}.
\label{AfidelityPE}
\end{equation}
This fidelity is always  smaller than the optimal universal cloning fidelity $F_{opt}^{1\rightarrow 2}$. In the  case of qubits the average fidelity of telecloning using  a partially entangled state reduces to 
\begin{equation}
F^{1\rightarrow 2, PE}_{d=2}=\frac{N_{\theta}^{2}}{12}\left[5|a|^{4}c_1^{2}+5|b|^{4}c_2^{2}+|a|^{2}|b|^{2}(1+8c_1c_2)\right],
\end{equation}
where 
\begin{equation}
N_{\theta}=\frac{\sqrt{2}}{\sqrt{a^{2}c_1^{2}+b^{2}c_2^{2}}}.
\end{equation}
As we can see from Eq. (\ref{AfidelityPE}), the lack of maximal entanglement not only reduces the fidelity of the broadcast clones but also makes the cloning process  state-dependent. As indicated by Eq. (\ref{RhoreducedPE}), this is due to the fact that the coefficients of the channel are introduced in the state of the clones. It is possible to improve the fidelities of the copies, even to optimal fidelity by using  additional steps. In the following sections we distinguish between the cases where the channel can be described by a state with maximal Schmidt rank or cases in which it may not be described in this fashion.

\section{Quantum channel of maximal Schmidt rank}
\label{SECTION-IV}

In the previous section we have shown that the use of a partially entangled channel leads to a telecloning process characterised by a suboptimal state-dependent fidelity. Here, we will show that it is possible to achieve the optimal state-independent fidelity at the expense of making the telecloning process probabilistic. 

This approach  allows the coefficients of the state to be cloned to be decouple from those of the channel. We can cast the state of the total system, Eq. (\ref{sys1}), before the telecloning process, as
\begin{equation}
\ket{\psi}_{XPAC} =  \frac{1}{d}\sum\limits\limits_{n,m=0}^{d-1}\ket{\tilde{\Phi}_{nm}}_{XP}\otimes U_{nm}^{-1}\sum\limits\limits_{j=0}^{d-1}\alpha_{j}\ket{\phi_{j}}_{AC},\label{eq:con estados sim}
\end{equation}
where the states $\{\ket{\tilde{\Phi}_{nm}}\}$ are defined by
\begin{equation}
\ket{\tilde{\Phi}_{nm}}_{XP}=\sum\limits\limits_{k=0}^{d-1}c_{k}\omega^{kn}\ket{k\ominus m}_{X}\ket{k}_{P}\label{no-distinguibles},
\end{equation}
with $n\ominus m=(n-m)~mod(d)$, and
\begin{equation}
U_{nm}^{-1}\sum\limits\limits_{j=0}^{d-1}\alpha_{j}\ket{\phi_{j}}_{AC}=\sum\limits\limits_{j=0}^{d-1}\omega^{-(j+m)n}\alpha_{j}\ket{\phi_{j\oplus m}}_{AC}
.\end{equation}
The coefficients of the quantum channel are now transferred into the new set $\{\ket{\tilde{\Phi}_{nm}}\}$ of generalised Bell states. These are non-orthogonal and cannot be perfectly distinguished in a deterministic fashion. Therefore, the selection of the correct unitary $U_{nm}$ is ambiguous and leads to errors in the clones. 

In order to avoid this limitation and improve the performance of the telecloning process, we can use probabilistic processes such as quantum state discrimination \cite{Chefles-DIS,review-dis} or quantum state separation \cite{Chefles-SEP}. These processes allow the correct reconstruction operation $U_{nm}$ to be chosen, or to improve the probability of choosing  the correct reconstruction operation in a controlled way. It is also possible to use minimum error discrimination \cite{Helstrom,Holevo-ME,Yuen-Me}, which is a deterministic discrimination strategy. In subsections \ref{USD}, \ref{ME} and \ref{QSS}  we analyse the performance of the telecloning process when  combined with each of these discrimination strategies.

\subsection{Telecloning combined with unambiguous state discrimination}
\label{USD}

The set $\{\ket{\tilde{\Phi}_{nm}}\}$ of $d^2$ non-orthogonal, partially entangled states can be transformed into a set of separable states. This is done by applying the $GXOR$ gate onto the $XP$ system. This gate is defined as
\begin{equation}
GXOR_{PX}\ket{n}_{P}\ket{m}_{X}=\ket{n}_{P}\ket{n\ominus m}_{X}.
\end{equation}
The GXOR gate is a generalisation of $CNOT$ for qudits and it can entangle and disentangle two qudit states. After applying this gate the state given by Eq. (\ref{eq:con estados sim}) is transformed into the state 
\begin{align}
 GXOR_{PX}\ket{\psi}_{PXAC} = & \frac{1}{d}\sum\limits\limits_{n,m=0}^{d-1}\sum\limits\limits_{k=0}^{d-1}c_{k}\omega^{kn}\ket{k}_{P}\ket{m}_{X}\nonumber \\
  & \otimes U_{nm}^{-1}\sum\limits\limits_{j=0}^{d-1}\alpha_{j}\ket{\phi_{j}}_{AC},
\end{align}
where the systems $X$ and $P$ are factorizable. The $X$ system is then projected into its computational basis by a  projective measurement.  The $X$ system is projected onto the state $\ket{m}_X$ with probability 
\begin{equation}
P_m=\sum\limits\limits_{k=0}^{d-1}\left|c_{k+m}\right|^{2}\left|\alpha_{k}\right|^{2}.
\end{equation}
The state of the total system, which depends on the results of the measurement on system $X$, becomes
\begin{align}
\label{despx}
\ket{\psi_{m}}_{XPAC} =& \frac{1}{\sqrt{P_{m}}}\frac{1}{d}\sum\limits\limits_{n=0}^{d-1}\ket{\Psi_{n}}_{P}\ket{m}_{X}\\ \nonumber
&\otimes\sum\limits\limits_{j=0}^{d-1}\omega^{-(j+m)n}\alpha_{j}\ket{\phi_{j\oplus m}}_{AC},
\end{align}
where the states
\begin{equation}
\ket{\Psi_{n}}_{P} =  \sum\limits\limits_{k=0}^{d-1}c_{k}\omega^{nk}\ket{k}_{P},\quad n=0,\ldots,d-1 
\label{simetricos}
\end{equation}
are a set of $d$ symmetric states. These are symmetric because  they are defined by the successive action of the operator $Z=\sum_{j=0}^{d-1}\omega^{j}\ket{j}\bra{j}$ over the seed state $\ket{\Psi_0}_P=\sum_{k=0}^{d-1}c_k\ket{k}_P$, namely $\ket{\Psi_n}_P=Z^n\ket{\Psi_0}_P$. Thus, the problem of discriminating among the $d^2$ states $\ket{\tilde{\Phi}_{nm}}$ of the bipartite system $XP$ is reduced to distinguishing among the $d$ states $\ket{\Psi_n}_P$ of the system $P$. These states are also mutually non-orthogonal and so they cannot be  deterministically distinguishable. We will now apply unambiguous state discrimination (USD) to the set $\{\ket{\Psi_n}_P\}$. It has been demonstrated that this discrimination scheme exist only  for sets of linearly independent states \cite{Chefles-USD}. In our case this corresponds to a set of non-vanishing $c_j$ coefficients, or equivalently, to a quantum channel of full Schmidt rank.

Now we apply the USD protocol to states $\ket{\Psi_n}_P$. This scheme corresponds to a unitary transformation $U^{USD}_{PX}$ acting on the bipartite system $PX$ followed by a von Neumann measurement on system $X$ \cite{Chefles-USD-SYM}. The unitary transformation $U^{USD}_{PX}$ is given by
\begin{equation}
U^{USD}_{PX}\ket{\Psi_{n}}_{P}\ket{m}_{X}=\sqrt{p_{d}}\ket{u_{n}}_{P}\ket{m}_{X}+\sqrt{1-p_{d}}\ket{\chi_{n}}_{P}\ket{m\oplus1}_{X},\label{eq:discrimination}
\end{equation}
where the set $\{ \ket{u_l}_P \}$ is composed by $d$ mutually orthogonal and distinguishable states of system $P$, the states $\{\ket{\chi}_P\}$ are $d$ linearly dependent states and $p_d$ is the  probability of successful discrimination of the states $\ket{\Psi_n}$. All the states $\ket{\Psi_n}$ are generated with the same  probability $1/d$, so the discrimination probability is the same for each of them, and is given by $p_d=d |c_{min}| ^2$ where $c_{min}$ is the channel coefficient with the smallest absolute value. The state $\ket{m\oplus 1}$ is orthogonal to the state $\ket{m}$, thus it is possible to know when the discrimination process is successful by means of a projective measurement on system $X$. 
The states $\{\ket{\chi_l}\}$ are given by
\begin{equation}
\ket{\chi_{l}}=\frac{1}{\sqrt{d}}\frac{1}{\sqrt{1-p_d}}\sum\limits\limits_{m,n=0}^{d-1}\omega^{(l-m)n}\sqrt{c_{n}^{2}-c_{min}^{2}}\ket{m}.
\end{equation}
These $d$ states belong to a subspace of dimension $d-1$ and are thus linearly dependent, which forbids the possibility of applying a further stage of unambiguous discrimination. The states $\{\ket{u_k}\}$ turn out to be the Fourier transform of the computational basis \cite{DelgadoC}, that is
\begin{equation}
\ket{u_{n}}=\mathcal{F}\ket{n}=\frac{1}{\sqrt{d}}\sum\limits\limits_{k=0}^{d-1}\omega^{kn}\ket{k}.
\end{equation}
Therefore, the application of the inverse Fourier transform on the system $P$ allow the states $\ket{\Psi_n}$ to be discriminated in the computational basis. First applying the discrimination unitary operation $U$ and then the inverse Fourier transform, we find that the state of the joint system $PXAC$ is given by
\begin{align}
&\ket{\tilde{\psi}_m}=\frac{\sqrt{p_{d}}}{d\sqrt{P_{m}}}\left(\frac{1}{\sqrt{d}}\sum\limits\limits_{n=0}^{d-1}\ket{n}_{P}U_{nm}^{-1}\sum\limits\limits_{j=0}^{d-1}\alpha_{j}\ket{\phi_{j}}_{AC}\right)\ket{m}_{X}&
\nonumber\\ 
 &+\frac{\sqrt{1-p_{d}}}{d\sqrt{P_{m}}}\left(\sum\limits\limits_{n=0}^{d-1}\mathcal{F}_{P}^{-1}\ket{\chi_{n}}_{P} U_{nm}^{-1}\sum\limits\limits_{j=0}^{d-1}\alpha_{j}\ket{\phi_{j}}_{AC}\right)\ket{m\oplus 1}_{X},&\label{USS}
\end{align}
where we have defined $\ket{\tilde{\psi}_m}=\mathcal{F}_{X}^{-1}U_{PX}^{USD}\ket{\psi_{m}}_{XPAC} $. We now perform a projective measurement on system $X$. If the outcome of the measurement is the eigenvalue associated with the state $\ket{m}_X$, which occurs with probability $p_d$, then it is possible to conclusively discriminate  the states of system $P$. We thereby obtain the values of $n$ and $m$. These are transmitted by classical communications to the carriers of systems $A$ and $C$, who separately apply  the local reconstruction unitary operations $U_{nm}^A$ on each ancilla system and $U_{nm}^C$ on each system encoding a clone. These reconstruction operations are given by
\begin{equation}
U_{n,m}^{A}  =  \sum\limits\limits_{j=0}^{d-1}\omega^{-(j+m)n}\ket{j}\otimes\bra{j\oplus m}\label{UA}
\end{equation}
and
\begin{equation}
U_{nm}^{C}  =\sum\limits\limits_{j=0}^{d-1}\omega^{(j+m)n}\ket{j}\otimes\bra{j\oplus m}. \label{UC}
\end{equation}
After these transformations the copies encoded in the systems $C$ are mixed states with optimal fidelity $F_{opt}^{1\rightarrow M}$ with respect to the input qudit state. Thus, the combination of  telecloning process via a partially entangled pure state and an unambiguous discrimination stage leads to a probabilistic telecloning process which produces optimal copies with probability $p_d=d |c_{min}|^2$ for any values of $M$ and $d$. 

In the case that the discrimination process fails, with probability $1-p_d$, the result of the measurement in the $X$ system turns to be $m\oplus 1$. The fidelity of the clones turns out to be non-optimal and depends on the state of system $X$ after the first measurement (through $m$) and on the input state. Indeed, if the discrimination fails, the state of the $AC$ system correspond to the second term of Eq. (\ref{USS}), namely
\begin{equation}
\ket{\tilde{\psi}_m}^{fail} =\frac{1}{d\sqrt{P_{m}}}\left(\sum\limits\limits_{n=0}^{d-1}\mathcal{F}_{P}^{-1}\ket{\chi_{n}}_{P} U_{nm}^{-1}\sum\limits\limits_{j=0}^{d-1}\alpha_{j}\ket{\phi_{j}}_{AC}\right)\ket{m\oplus 1}_{X}\label{USSF}.
\end{equation}
Starting from Eq. (\ref{USSF}), we can construct the density matrix for the final state in the case of a failed discrimination attempt, and then obtain the reduced density matrix of one of the copies, namely $C_2$, by calculating, in the case of the $1\rightarrow 2$ telecloning,
\begin{equation}
\rho_{n,m}^{C_{2},fail}=Tr_{C_{1}}\left[Tr_{PA}\left[|\tilde\psi_{m}\rangle^{fail}\langle\tilde\psi_{m}|\right]\right].
\end{equation}
The fidelity of the clones  $F_{m}^{1\rightarrow2,fail}=\langle\psi|\rho_{m}^{C_{2},fail}|\psi\rangle$  is given by
\begin{align}
F_{m}^{1\rightarrow 2, fail}=&\frac{1}{2(d+1)}+\frac{1}{2(d+1)}\frac{1}{{P}_{m}}\sum\limits\limits_{j=0}^{d-1}\left|\alpha_{j+m}\right|^{2}\left|\alpha_{j}\right|^{2} \label{hola}\\ \nonumber 
&\times \left\{ (2+d)\left(c_{j+m}^{2}-c_{min}^{2}\right)\right\}.
\end{align}
This depends on the result of the first measurement, given by the value of $m$, and on the state to be cloned, so that, in the case of failure, the process is neither universal nor optimal.
The average failure fidelity is computed by integrating over all the possible input states and summing over all the possible outcomes of the first measurement on system $X$, 
\begin{equation}
F^{fail}  = \int d\alpha\sum\limits\limits_{m=0}^{d-1}P_{m}F_{m}^{1\rightarrow 2, fail}, \label{failurs}
\end{equation}
where $\int d\alpha$ represents integration over all pure $d$-dimensional qudits. Using the identity $\int d\psi\left|\psi_{j}\right|^{2}\left|\psi_{k}\right|^{2}=(\delta_{j,k}+1)/d(d+1)$ and Eq. (\ref{hola}) we obtain
\begin{equation}
F^{fail} =\frac{1}{d}. 
\end{equation}
A simple cloning strategy consists of estimating the state to be cloned and then creating as many copies of the estimated state as needed. The fidelity of this process is given by the optimal state estimation fidelity $F_{est}=2/(d+1)$. Thus, the average telecloning fidelity obtained in cases which the discrimination attempts fail is always smaller than that obtained when estimating.

Finally, as a measure of the quality of the total process obtained by concatenating telecloning to unambiguous state discrimination, it is possible to compute the total average fidelity $F_{av}$, which includes failures and successes in the discrimination stage. This average fidelity is obtained by adding the optimal fidelity of cloning  $F_{opt}^{1\rightarrow M}$ weighted by the optimal discrimination probability $p_d$ and the average fidelity $F^{fail}$, given in Eq. (\ref{failurs}), weighted by the failure probability $1-p_d$:
\begin{equation}
F_{av}=F_{opt}^{1\rightarrow M} p_d+F^{(fail)}(1-p_d).
\end{equation}
In the case of $1\rightarrow 2$ telecloning of qudits, this fidelity is given by
\begin{equation}
F_{av}^{1\rightarrow 2} = p_{d}(\frac{3+d}{2+2d})+(1-p_{d})\frac{1}{d},
\end{equation}
which  depends on the quantum channel through $p_d$. We can compare this fidelity with the classical fidelity $F_{set}$ for $1\rightarrow2$ cloning processes, i.e., the fidelity of
the optimal measure-and-prepare cloner. For  the $1\rightarrow 2$ telecloning of qudits we look for dimensions where
\begin{equation}
p_{d}(\frac{3+d}{2+2d})+(1-p_{d})\frac{1}{d}\ge\frac{2}{d+1}
\end{equation}
holds. In addition, the condition $p_d \le 1$ also holds. This implies that $|c_{min}|^2\le 1/d$. Therefore, it is  possible to outperform the classical cloning fidelity in every dimension $d$ if the condition for the channel 
\begin{equation}
|c_{min}|^2 \ge \frac{2}{d(d+2)},
\end{equation} 
is fulfilled. Hence, for certain quantum channels, even when including  clones produced in cases with failed discrimination processes, it is possible to achieve higher telecloning fidelities than in the classical case.

\subsection{Telecloning combined with minimum error discrimination}
\label{ME}

Minimum error discrimination of quantum states \cite{Helstrom,Holevo-ME,Yuen-Me,Bae-ME} is based on the minimisation of the average error  when making guesses about a set of states. Operationally, following the same procedure as using unambiguous state discrimination, and after a projective measurement in the system $X$, the state of the $PAC$ system is given by
Eq. (\ref{despx}),
\begin{equation}
\ket{\psi_{m}}=\frac{1}{\sqrt{P_{m}}}\frac{1}{d}\sum_{n=0}^{d-1}\ket{\Psi_{n}}_{P}\sum_{j=0}^{d-1}\omega^{-n(j+m)}\alpha_{j}\ket{\phi_{j\oplus m}}_{AC},
\end{equation}
Now, in order to discriminate between  the $|\Psi_n\rangle$ states we apply the inverse Fourier transform over $P$,
\begin{equation}
\mathcal{F}^{-1}\ket{\Psi_{n}}_{P}  =  \frac{1}{\sqrt{d}}\sum_{k=0}^{d-1}\sum_{j=0}^{d-1}c_{k}\omega^{k(n-j)}\ket{j},
\end{equation}
and subsequently we projectively  measure in the canonical basis. If the result is the one associated with the state $|n\rangle$, the state of the system is given by
\begin{equation}
_{P}\bra{n}\mathcal{F}_{P}^{-1}\ket{\psi_{m}}  =  \frac{1}{\sqrt{P_{m}}}\frac{1}{\sqrt{d}}\sum_{j=0}^{d-1}c_{j+m}\alpha_{j}\omega^{-n(j+m)}\ket{\phi_{j\oplus m}}_{AC},
 \end{equation}
which is non-normalized, but with norm  $\tilde P_n$. Knowing the values of $n$ and $m$ we apply the reconstruction operations given by Eqs. (\ref{UA}) and (\ref{UC}). The new state of the $AC$ system is given by
\begin{equation}
\ket{\psi_{nm}}  =  \frac{1}{\sqrt{\tilde{P}_{n}}}\frac{1}{\sqrt{P_{m}}}\frac{1}{\sqrt{d}}\sum_{j=0}^{d-1}c_{j+m}\alpha_{j}\omega^{-nm}\ket{\phi_{j}}_{AC}\label{gr}.
\end{equation}
The probability of measuring the state associated with $n'$ is given by 
$
\tilde{P}_{n}  =  \frac{1}{P_{m}}\frac{1}{d}\sum_{j=0}^{d-1}\left|c_{j+m}\right|^{2}\left|\alpha_{j}\right|^{2}
$, and replacing Eq. (\ref{pm}), we get $\tilde{P}_{n}=\frac{1}{d}$.

As in the previous cases, it is possible to compute the fidelity of the copies in the $1 \rightarrow 2$ telecloning using Eq. (\ref{1-2}) and Eq. (\ref{gr}) to get first the density matrix of the $AC_1C_2$ system, and then the reduced density matrix of one of the copies, namely $C_2$, which is given by
\begin{equation}
\rho_{n,m}^{C_{2}}=Tr_{C_{1}}\left[Tr_{A}\left[|\psi_{n,m}\rangle_{A,C_{1},C_{2}}\langle\psi_{n,m}|\right]\right].
\end{equation}
Hence, the fidelity of the copies, which is given by $F_{m}^{1\rightarrow2,ME}=\langle\psi|\rho_{n,m}^{C_{2}}|\psi\rangle$, has the explicit form
\begin{equation}
 F_{m}^{1\rightarrow2,ME} =  \frac{1}{2(d+1)}\frac{1}{d^{3}}+\frac{(2+d)}{2(d+1)}\frac{1}{P_{m}}\frac{1}{d^{3}}\left(\sum_{j=0}^{d-1}|\alpha_{j}|^{2}c_{j+m}\right)^{2}.
\end{equation}
This fidelity depends on the result of the first measurement and this dependence is carried in the $m$ indexes and in the input state, so that, after applying minimum error discrimination the process is no longer universal, but still symmetric. Finally, the overall fidelity of the process is calculated by the weighted sum of all the possible measurements results, i.e. $\langle F^{ME}\rangle=\sum_{n,m=0}^{d-1}\tilde{P}_{n}P_{m}F_{nm}^{2}$, which is given explicitly by
\begin{equation}
\langle F^{ME}\rangle  =  \frac{1}{2(d+1)}\frac{1}{d^{3}}+\frac{(2+d)}{2(d+1)}\frac{1}{d^{3}}\sum_{m=0}^{d-1}\left(\sum_{j=0}^{d-1}|\alpha_{j}|^{2}c_{j+m}\right)^{2}.
\end{equation}
The last expressions corresponds to $F^{1 \rightarrow 2 \text{,PE}}/d^3$, which  means that the overall fidelity of the process with minimum error discrimination is always worse than the process without any discrimination protocol. In the case of qubits this fidelity is given by 

\begin{equation}
\langle F^{ME}\rangle^{1\rightarrow 2}_{d=2}=\frac{N_{\theta}^{2}}{96}\left[5|a|^{4}c_1^{2}+5|b|^{4}c_2^{2}+|a|^{2}|b|^{2}(1+8c_1c_2)\right].
\end{equation}

\subsection{Telecloning combined with quantum state separation}
\label{QSS}

The process of quantum state separation \cite{Chefles-SEP} consist of changing the separation  between different states by a different and, in general, larger one. This separation is given by the inner product of the corresponding states. Therefore, the increase of this value between non-orthogonal states allows us to increase the chance of discrimination between them. This process is characterised by a unitary operation $S$ and is a probabilistic process. In the case of a set of equidistant symmetrical states, as in the case of the states given by Eq.(\ref{simetricos}), and considering that we want to symmetrically increase the separation of these states, the probability of this process is given by $p_{SEP}=c_{min}^2/\tilde{c}_{min}^2$, where $\tilde{c}_{min}$ is the minimal coefficient that we want to achieve.
If we set the increase of the distance between the symmetrical states on Eq. (\ref{simetricos}) so that after applying the suitable quantum state separation operation over system $P$ they will be mutually orthogonal,  it is then possible to discriminate and correctly choose the unitary reconstruction operations and at the end have optimal copies of the input state, i.e., with fidelity $F_{opt}^{1\rightarrow 2}$, but with a limited probability $p_{SEP}^{ort}=c_{min}^2/d$, which is always smaller than the probability achieved by unambiguous state discrimination.

However, it is also  possible  to change the separation of this state arbitrarily, not necessarily   to orthogonality. If we apply  this procedure  over the system $P$,  considering the states of Eq. (\ref{simetricos}), so that the new set of coefficients are denoted by $\tilde{c}_k$, and without other processes of discrimination, it is possible improve the fidelity of the copies, which in the case of $1\rightarrow 2$ telecloning will be given  by 
\begin{equation}
F_{m}^{1\rightarrow 2,SEP}=\frac{1}{2(d+1)}+\frac{(2+d)}{2(d+1)}\frac{1}{\tilde{P}_{m}}\left(\sum_{k=0}^{d-1}|\alpha_{k}|^{2}\tilde{c}_{k+m}\right)^{2},
\end{equation}
where $\tilde{P}_{m}=\sum_{j=0}^{d-1}|\alpha_{j}|^{2}|\tilde{c}_{j+m}|^{2}$. The probability of obtain this fidelity is given by $p_{SEP}$.
The weighted fidelity over every possible outcome $m$ is given by
\begin{equation}
\langle F^{SEP}\rangle^{1\rightarrow 2}= \frac{1}{2(d+1)}+\frac{(2+d)}{2(d+1)}\sum\limits_{m=0}^{d-1}\left(\sum_{k=0}^{d-1}|\alpha_{k}|^{2}\tilde{c}_{k+m}\right)^{2}
\end{equation}
In the case of telecloning of qubits, this quantity reduces to
\begin{equation}
\langle F^{SEP}\rangle^{1\rightarrow 2}_{d=2}=\frac{\tilde{N}_{\theta}^{2}}{12}\left[5|a|^{4}\tilde{c}_1^{2}+5|b|^{4}\tilde{c}_2^{2}+|a|^{2}|b|^{2}(1+8\tilde{c}_1\tilde{c}_2)\right]
\end{equation}
where
\begin{equation}
\tilde{N}_{\theta}=\frac{\sqrt{2}}{\sqrt{a^{2}\tilde{c}_1^{2}+b^{2}\tilde{c}_2^{2}}}.
\end{equation} 
If we use the quantum state separation process to orthogonalize in the qubit case, the fidelity achieved is the optimal 5/6 for any input state and the probability of success  is $c^2_{min}/2$, which is four times smaller than the probability of achieving the same fidelity when using unambiguous state discrimination.

\section{Quantum channel of non-maximal Schmidt rank}
\label{SECTION-V}
In the case where $c_j=0$, for some $j$, the states of  Eq. (\ref{simetricos}) are linearly dependent and it is not possible to unambiguously  distinguish between them, so that it is necessary to resort to other discrimination schemes. A useful discrimination scheme in this scenario is  maximum confidence state discrimination (MCD) \cite{Croke-MC}. This scheme allows us to discriminate between linearly dependent states but with errors associated  with the identification  of some of them, even if we permit non-conclusive  results in the discrimination measurements. Therefore, it is possible to construct a set of measurements which enables us to identify the states with the maximum possible confidence.

We apply, as in the USD case, the transformation of maximum confidence discrimination on the system $X$  using  the system $P$ as ancilla system.  The initial state of the ancilla system is $\ket{m}$. If this state remains unchanged, the discrimination process succeeds  and it is possible discriminate with maximum confidence. In any other case, the result is inconclusive. After applying the MCD unitary transformation $U^{MC}$, the state of the system is given by
\begin{align}
\ket{\tilde{\tilde{\psi}}}=&\sqrt{1-p_{?}}\ket{m}_{X}\sum\limits\limits_{n=0}^{d-1}\ket{\tilde{u}_{n}}_{P}U_{nm}^{-1} \sum\limits\limits_{j=0}^{d-1}\alpha_{j}\ket{\phi_{j}}_{AC} \nonumber\\
  & +\sqrt{p_{?}}\ket{m\oplus1}_{X}\sum\limits\limits_{n=0}^{d-1}\ket{\tilde{\chi}_{n}}_{P}U_{nm}^{-1}\sum\limits\limits_{j=0}^{d-1}\alpha_{j}\ket{\phi_{j}}_{AC},
\end{align}
where $\ket{\tilde{\tilde{\psi}}}=U_{XP}^{MC}\ket{\psi_{m}}_{XPAC} $. The states $\ket{\tilde{u}}_n$ are orthonormal states given by
\begin{equation}
\ket{\tilde{u}_{n}} =  \frac{1}{\sqrt{N}}\sum\limits_{k=0}^{N-1}e^{2\pi kn/d}\ket{k}_{2},
\end{equation}
and the states $\ket{\tilde{\chi}}$ are normalized  non-orthogonal states, given by
\begin{equation}
\ket{\tilde{\chi}_{n}} =\sum\limits_{k=0}^{N-1}\sqrt{\frac{c_{k}^{2}-c_{min}^{2}}{p_{?}}}e^{2\pi kn/d}\ket{k}.
\end{equation}
$N$ is the number of non-vanishing coefficients $c_j$. As in the USD case, if we get the outcome $m$ by measuring the system $X$, we will be able to perform an inverse Fourier transform and measure the system $P$ on the computational basis, obtaining the outcome $n$, and then perform the transformations of Eq. (\ref{UA}) and (\ref{UC}). The confidence of the process, i.e., the probability  of  correctly distinguish  the state $n$ through  the  measurement outcome  $n$ is $N/d$. Additionally, the probability of getting a non-conclusive outcome is $p_?=1-Nc_{min}^2$, where in this case $c_{min}$ is the  non-vanishing coefficient with the smallest modulus.

\section{Summary}
\label{SECTION-VI}
We presented a general scheme to probabilistically teleclone qudit states via partially pure entangled channels as described in Eq. (\ref{partialst}). After introducing  these kind of channels, the process is still symmetric but no longer universal. However, it is still possible to recover the universality of the cloning process in the case of maximally ranked Schmidt channels, by using unambiguous state discrimination and  turning the process into a probabilistic one.  The same result could be achieved using quantum state separation but with a much smaller probability.  If instead of using this processes we use minimum error discrimination to try to improve the results, the overall fidelities are even worse than not using any strategy.
In \cite{Li} the authors propose a similar probabilistic scheme but just in the  $1\rightarrow 2$ case and  maximal Schmidt rank using a \textit{pseudo control unitary operation}  which corresponds to a special case of our unambiguous state discrimination strategy.
\begin{acknowledgements}
This work was supported by Millenium Scientific Initiative Grant No. RC130001. N. C. and G. A. acknowledge support from CONICyT. We thank M. W. van Mourik and M. G. Higgins for their valuable contribution to the improvement of the manuscript.  
\end{acknowledgements}



\end{document}